# Bringing the Interaction of Silver Nanoparticles with Bacteria to Light

Simone Normani[1], Nicholas Dalla Vedova[1], Guglielmo Lanzani[1,2], Francesco Scotognella[2] and Giuseppe Maria Paternò[1]*

[1]Center for Nano Science and Technology@PoliMi, Istituto Italiano di Tecnologia, Via Giovanni Pascoli, 70/3, 20133 Milano, Italy

[1]Physics Department, Politecnico di Milano, Piazza Leonardo Da Vinci, 32, 20133 Milano, Italy

*Authors to whom correspondence should be addressed: giuseppe.paterno@iit.it

**ABSTRACT**

In the last decades the exploitation of silver nanoparticles in novel antibacterial and detection devices have risen to prominence for their well-known specific interaction with bacteria. The vast majority of studies focus on the investigation over the mechanism of action underpinning bacterial eradication, while little efforts have been devoted to the modification of silver optical properties upon interaction with bacteria. Specifically, given the characteristic localized surface plasmon resonance of silver nanostructures, which is sensitive to changes in the charge carrier density or in the dielectric environment, these systems can offer a handle in the detection of bacteria pathogens. In this review, we present the state of art of the research activity on the interaction of silver nanoparticles with bacteria, with emphasis on the modification of their optical properties. This may indeed lead to easy color reading of bacterial tests and pave the way to the development of nanotechnology silver based bacterial detection.

**INTRODUCTION**

One of the great advances in medical science was the discovery of antibiotics, which provided humankind with an effective tool to fight off bacterial infections. However, for how effective a solution this was in fighting pathogenic microbes, in recent years the extensive use of antibiotics has led to bacteria developing resistance to drug therapy and the emergence of multidrug-resistance (MDR) in mutant species, which poses an obvious threat to human welfare[1–3]. Hence the need arises for researching alternatives to antibiotics, which will overcome the rise of drug-resistant microbes: in this context, the potential of metals, and in particular silver, as an option for antibacterial devices has been recognized. Historically, silver has been widely regarded for its antimicrobial properties[4]. Although much remains to be understood about the mechanism that contribute to its bactericidal action, advancements in research have made it possible to study more closely the interaction between the metal and the pathogens, in order to obtain a clearer picture of the underlying biochemical mechanics. In particular, recent advances on nanotechnology have piqued the interest of researchers into the interaction between silver nanoparticles (AgNPs) and bacteria. This may lead to useful applications in the medical field, for instance in rendering metallic surgical implants safer by inhibiting pathogen growth[5], but also in surface and device sanitization, and in the design and fabrication of bio-detectors for environmental and food-related monitoring[6,7]. Many studies concerning AgNPs pointed out the relative high surface-to-volume ratio, which is granting a large effective area of interaction with the pathogens per unit mass. This in turn leads to larger inhibition of the bacteria virulence when using nanoparticles as opposed to a flat metal surface. Similar effect has been observed on pathogens such as *Escherichia coli* on nano-rough gold surfaces, which were shown to prevent the development of fimbriae and possibly damage the cell membrane and its metabolism[8].

Although silver bactericidal effect has been studied for several decades, the modulation of its optical properties upon interaction with bacteria has been largely overlooked. This is a pity, since investigations over these effects can permit to gain more insight into the biophysics underpinning their interaction, as well as permitting to develop novel colorimetric sensing devices based on the optical changes of the medium. While not yet thoroughly understood, the changes in optical properties of AgNPs when interacting with bacteria can indeed be used as a key parameter to monitor the presence of pathogens. For example, the presence of bacteria can be monitored via the localized surface plasmon resonance shift, which can originate from the change in the dielectric environment and increase in free electron density, *i.e.* due to structural modifications and/or as a consequence of silver oxidative dissolution. In this review, after summarizing briefly the antibacterial mechanism of action of AgNPs, we will give an overview on the possibility to use these systems to reveal optically the presence of bacteria.

**MECHANISMS OF INTERACTION OF SILVER NANOPARTICLES WITH BACTERIA**

Though the complex chain of events leading to bacterial eradication is still being investigated in details, the main antibacterial agent of AgNPs has been identified in the $Ag^+$ species, while the nanoparticles themselves serve the function of silver atom reservoirs[9]. $Ag^+$ forms as the result of an oxidative reaction from neutral $Ag^0$, which requires the presence of an oxidizer. Released $Ag^+$ ions can then react with common organic substances, such as thiols, as well as organic amines and phosphates, with the Ag-S bond being the most common[10]. This is the main mechanism that allows the positive ions to bind to the cell membrane of bacteria, which drives the process of membrane decay and cell death. The role of oxygen and oxidative processes was specifically investigated by Xiu *et al.*[11], who compared the antibacterial cytotoxicity of AgNPs in anaerobic and aerobic environments. The paper showed negligible effects of the AgNPs on the bacterial growth rate in the anaerobic case, in spite of the high particle concentration, largely above the estimated lethal concentration of $Ag^+$ ions. This observation leads to conclude that the presence of oxygen is necessary for the release of $Ag^+$ ions, confirming that the bacterial toxicity of AgNPs stems from the oxidative dissolution process. A detailed description of the $Ag^+$ release mechanisms was already proposed in 2010[12], wherein one can in fact see that oxygen intermediates play a substantial role in extracting the positive ions from the particle surface.

The role of the electrostatic interaction between the positively charged silver ions and bacteria cell walls has been elucidated in the last couple of decades. In particular, several papers showed a greater growth inhibition in gram-negative than gram-positive bacteria[13,14]. This was attributed to the structure of their respective cell walls: whereas gram-negative bacteria have a layer of negatively charged and flexible lipopolysaccharides at the exterior, the cell wall in gram-positive bacteria is principally composed of a thick layer of zwitterionic and rigid peptidoglycan. The superior effectivity of $Ag^+$ in inhibiting growth in gram-negative bacteria has been thus ascribed to the higher mechanical flexibility and density of negative charges in this kind of bacteria than the gram-positive counterparts[13]. In these regards, the role of Ag/cell wall electrostatic interaction has been evidenced by the fact that positively-charged AgNPs showed a much lower minimum inhibitory concentration value against *E. coli* (two orders of magnitude) with respect to their neutral counterpart.

The uptake of $Ag^+$ ions occurs mostly by endocytosis, which has been argued to cause accumulation of silver in lysosomes as well as increases in the permeability of the cell membrane by peroxidation of membrane lipids, which also lead to cytoplasm leakage[15,16]. A study by Ansari *et al.*[17] showed a gradually increasing degree of aggregation of AgNPs in the *E. coli* cytoplasm and loss of cytoplasmic material through the membrane as a function of AgNPs concentration. In addition, the dissolution process of $Ag^+$ ions not only degrades the cell membrane, but also leads to the generation of reactive oxygen species (ROS), which are a known cause of apoptosis. The degradation of the membrane combined with the biocidal action of ROS is therefore the most commonly accepted description of the bactericidal action of silver surfaces and AgNPs in particular[16,18,19]. The electrostatic interaction of AgNPs with cell membrane is the most relevant mechanism at play for detection purposes, as the variations in both size and total electric charge of the metallic particles influence the localized surface plasmon resonance, and therefore the resulting optical and spectroscopic response.

**OPTICAL RESPONSE OF SILVER NANOPARTICLES UPON INTERACTION WITH BACTERIA**
In AgNPs, the collective oscillations of the metal electron density (plasma oscillations) are confined to nanoscale volume and quantized, giving rise to discrete and localized plasmon states, the so-called localized surface plasmon resonance (LSPR). The fact that such a distinctive absorption feature is extremely sensitive to change in the dielectric environment and charge carrier density, make them viable options for use in optical detection. For instance, our group has reported recently on the combined use of plasmonic nanostructures and photonic crystals for building up photo/electrochromic devices[20–23]. Specifically to AgNPs, the LSPR usually manifests as a broad absorption peak spanning from the UV-visible[24,25] to near-infrared range[26], depending on particle size distribution and shape[27]. The oxidative dissolution of silver, which occurs thanks to the synergic effect of oxygen and a nucleophilic ligand, is responsible for the release of the actual toxicant $Ag^+$. It is interesting noting that such a reaction is exploited commercially for the extraction of gold from metal ore by using cyanide as nucleophilic ligand (gold cyanidation).

The role of oxidative dissolution in influencing the optical properties of AgNPs has been investigated in details by Morgensen and Kneipp[28] in 2014 (**Fig. 1**). In particular, they observed an intrinsic blue shift of the plasmon resonance in the extinction spectra of AgNPs exposed to oxidizing agents over time, which was mainly attributed to the reduction in particle size. This was preceded by a red shift when using an oxidizing substance such as cysteamine, due to the change in the refractive index. However, it is noted that the same was not observed with other media such as cyanide, whose refractive index change is much smaller, nor with direct oxidants such as hydrogen peroxide. This study allowed disentangling between the optical effects due to change in the dielectric environment (cystamine-induced red shift), from those related to the modification of the charge carrier density (oxidative dissolution-induced blue shift). However, in this study such effects were not investigated in presence of actual bacterial cells.

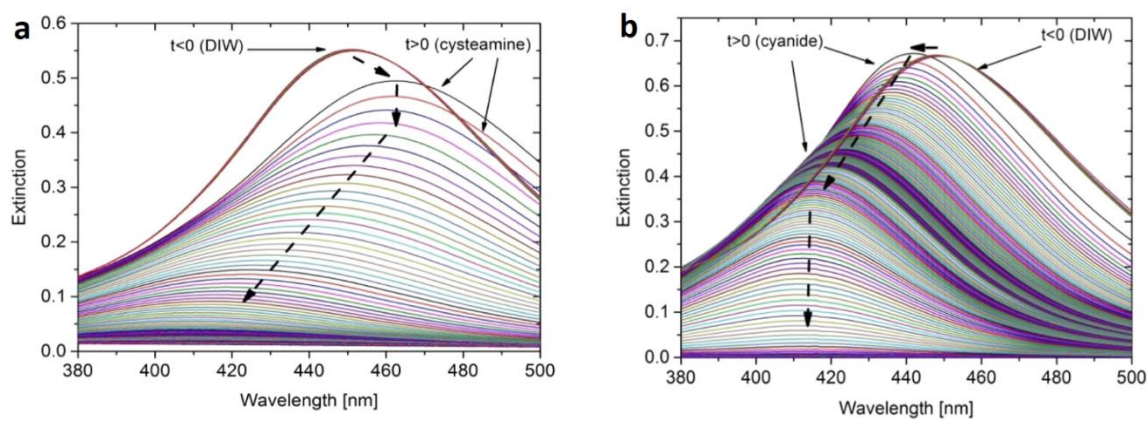

**Figure 1.** Time series of extinction spectra collected during dissolution of an AgNP film with 10 nm radius using (a) 1 mM cysteamine and (b) 0.1 mM cyanide. Spectra for $t < 0$ are measured in deionized water (DIW). Dashed arrows show the progression over typically 30−60 min. Adapted from Morgensen and Kneipp[28]. Copyright 2014, American Chemical Society.

The effect of *E. coli* exposure on the LSPR peak of AgNPs was reported by Sepunaru *et al.* in 2015[29]. Here, the authors observed essentially a strong attenuation of the absorption peak upon exposure to the physiological KCl buffer, which the authors connect to bacterial-induced severe particles aggregation (**Fig. 2**). On the other hand, the bacterial cellular environment caused a stabilization of the particles (**Fig. 2** inset) due to strong interaction between AgNPs and bacteria. Conversely, LSPR damping in presence of *E. coli* and *Pseudomonas aeruginosa* has been reported in a more recent paper[30]. This was attribute to the development of resistance to AgNPs, which stems from the production of the adhesive flagellum protein flagellin in swimming bacteria, which in turns leads to particles aggregation and reduction of their antibacterial activity. In general, these studies provide important insights that may be exploited for bacterial optical detection.

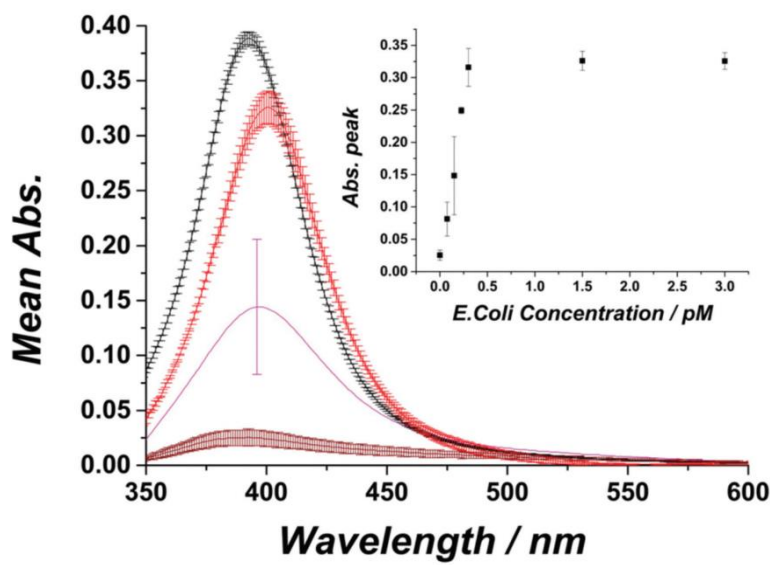

**Figure 2.** Absorption spectra of 10 nM AgNPs in water (black), in 0.1 M KCl solution (brown) and in 0.1 M KCl solution containing 0.3 pM (red) or 0.15 pM (pink) *E. coli* cells. (Inset) Titration curve of the absorption peak (at 399 nm) of 10 nM AgNPs in 0.1 M KCl solution containing different con- centrations of E. coli. Adapted from reference[29]. Copyright 2015. The Royal Society of Chemistry.

More recently, we have showed that the spectral position of the LSPR in silver nanoplates films undergoes a blue-shift upon exposure to bacteria (**Fig. 3**)[31,32]. In particular, exposure to the culture medium only (Luria-Bertani, LB) leads to a clear red-shift of the plasmon absorption, a result that can be attributed to the increase in the effective refractive index. On the other hand, a blue shift in the absorption band is observed after contamination with *E. coli*. This was ascribed to an increase of the charge carried density (bio doping), possibly owing to the oxidative dissolution process. Interestingly, contamination with the gram-positive *Micrococcus luteus* leads to a more convoluted effect, encompassing both a red shift due to the culture medium and an increase of the high-energy absorption (blue shift). Such an intricate read-out was related to a less sensitive interaction between the silver surface and gram-positive bacteria, due to the different cell wall structures. Integration of such a response with photonic crystals can be exploited to build up colorimetric devices for the detection of bacteria, in which the plasmon response is translated into a bacterial-induced color shift[31–33]. The enhanced interaction of silver with gram-negative bacteria suggests that AgNPs can be used not only as a standalone bactericidal and detection medium, but also as a way to functionalize other plasmonic surfaces. For instance, Yao *et al.* have reported on the functionalization of a gold surface with immobilized AgNPs[34]. leading to an enhancement of the plasmon resonance blue-shift as a function of the *E. coli* concentration. This can open the way for further options and approaches available for bacterial optical sensing.

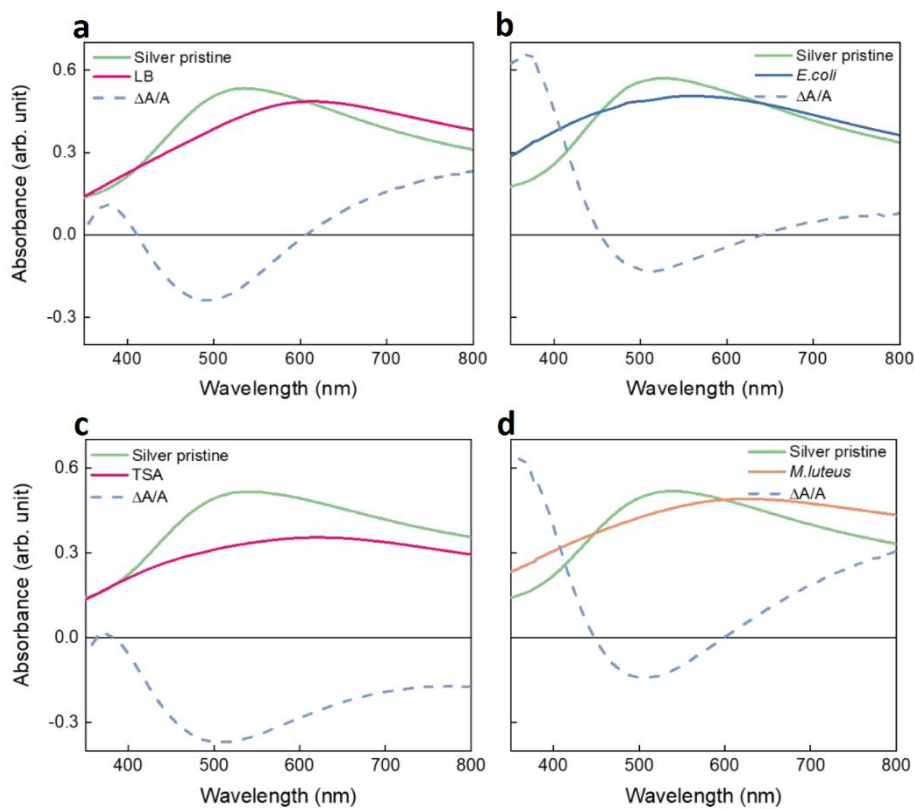

**Figure 3.** Optical absorption of 8 nm-thick silver layers composed of nanoplates on glass substrate before (silver pristine) and after exposition (a) to LB culture broth and (b) to *E. coli*, and before and after exposition (c) to the tryptic soy culture broth and (d) to *M. luteus*. The blue dashed lines represent the differential absorption expressed as ($Abs_{E.\ coli} - Abs_{\text{pristine}}$) / $Abs_{\text{pristine}}$. Adapted from reference[32]. Copyright Royal Society of Chemistry, 2020.

Conversely, it has been observed that the LSPR position and width of AgNPs can be taken as a descriptor for their antimicrobial efficacy. In particular, Mlalila et *al.* showed that smaller NPs exhibiting narrower LSRP perform better in terms of antimicrobial action than larger NPs with broader response.[35] In addition, they noted that positively-charged AgNPs showed an enhanced LSPR, a narrower width and a superior antimicrobial activity as compared to neutral ones[36]. The relatively high area-to-volume ratio of small NPs, as well as the more effective electrostatic interaction between small positive ions with the bacterial cell walls have been taken into account to explain those findings. This study can potentially provide useful information for the optimization of the material's microbe detection sensitivity.

## CONCLUSIONS

The research on the interaction of AgNPs with bacteria has produced very important results in the recent years, showing promising aspects that can be employed to overcome the issue of multi-drug resistance in pathogens. While a complete and accurate description of the chemical processes behind the AgNP cytotoxicity is still being discussed, the release of $Ag^+$ ion from the NP surface plays a leading role in facilitating the degradation of the cell membranes and bactericidal activity. At the same time, the high sensitivity of the LSPR position and width to the dielectric and chemical environment render them also useful optical detection tools. For instance, the oxidative dissolution process that is at the basis of the release of the $Ag^+$ toxicant upon interaction with bacteria, can lead to LSPR blue shift because of the increased charge carrier density. Furthermore, bacterial-induced particle aggregation causes clear damping of the LSPR, thus suggesting that also this parameter can be taken as a physical descriptor of the interaction. With the tools at our disposals, the fabrication of appropriately designed AgNPs for bacterial detection purposes is therefore possible, and may lead to the development of efficient monitoring and screening medical devices for various pathogens. However, the exact photophysics behind the oxidative dissolution of AgNPs and its spectroscopical signatures are not yet univocally clear: while the plasmon shift has been noticed in most works, both blue-shift or red-shift is reported, seemingly depending on the specific environment in consideration. There is therefore room for further research into the photophysics and specific mechanism of the AgNPs-bacteria interactions, in order to have precise knowledge of their optical response, and for potential applications of such devices as versatile and effective pathogen detectors.

## AUTHOR'S CONTRIBUTIONS

All authors contributed to manuscript drafting and revising, and figure creation. This article has been submitted to the journal Biophysics Reviews.

## ACKNOWLEDGMENTS

This work has been supported by Fondazione Cariplo, grant n° 2018-0979 and n° 2018-0505. F.S. thanks the European Research Council (ERC) under the European Union's Horizon 2020 research and innovation programme (grant agreement No. [816313]).

## CONFLICT OF INTERESTS

The authors declare financial support from the company "Dg for life". The authors clarify that the scientific conclusions that are presented in this study are not influenced by the activity of the company.

## DATA AVAILABILITY STATEMENT

Data sharing is not applicable to this article as no new data were created or analyzed in this study.